\documentclass[useAMS,usenatbib,usegraphicx]{mn2e}
\usepackage{fixltx2e}

\def\aap{A\& A}

\def\apj{ApJ}
\def\apjl{ApJL}

\def\mnras{MNRAS}

\def\aj{{AJ}}
\def\apjs{ApJS}
\def\mnras{{MNRAS}}

\def\pasp{{PASP}}

\def\'#1{\ifx#1i{\accent"13\i}\else{\accent"13#1}\fi}
\def\alamenos#1{$^{-#1}$}
\def\ala#1{$^{#1}$}

\def\diezala#1{10$^{#1}$}

\def\rhopdf{$\rho$-PDF}
\def\Npdf{$N$-PDF}
\def\VS{{V\'azquez-Semadeni}}

\def\BP{Ballesteros-Paredes}

\title[Column density PDFs]{Gravity or turbulence? II. Evolving column
density PDFs in molecular clouds}

 \author[Ballesteros-Paredes et al. ]
{ \parbox{7.0in}{Javier Ballesteros-Paredes\ala 1\thanks{e-mail:{\tt 
j.ballesteros@crya.unam.mx}}, 
Enrique V\'azquez-Semadeni\ala 1, 
Adriana Gazol\ala 1,
Lee W. Hartmann\ala 2, 
Fabian Heitsch\ala 3, and
Pedro Colin\ala 1
}\\
\\
\ala 1 Centro de Radioastronom\'ia y Astrof\'isica,
            Universidad Nacional Aut\'onoma de M\'exico, \\
            Apdo. Postal 72-3 (Xangari), Morelia,
            Michoc\'an 58089, M\'exico \\
\\
     \ala 2 Department of Astronomy, University of Michigan,  500
           Church Street, Ann Arbor, MI 48105, USA \\
\\
     \ala 3 Department of Physics and Astronomy, University of North
            Carolina Chapel Hill, \\ CB 3255, Phillips Hall, Chapel
            Hill, NC 27599, USA \\
}

\begin{document}


\date{Submitted to MNRAS, \today}

\pagerange{\pageref{firstpage}--\pageref{lastpage}} \pubyear{2006}

\maketitle

\label{firstpage}

\begin{abstract}

It has been recently shown that molecular clouds do not exhibit a
unique shape for the column density probability distribution function
(\Npdf). {Instead, clouds without star formation seem to possess a
  lognormal distribution, while clouds with active star formation
  develope a power-law tail at high column densities.  The lognormal
  behavior of the \Npdf\ has been interpreted in terms of turbulent
  motions dominating the dynamics of the clouds, while the power-law
  behavior occurs when the cloud is dominated by gravity.  In the
  present contribution we use thermally bi-stable numerical
  simulations of cloud formation and evolution to show that, indeed,
  these two regimes can be understood in terms of the formation and
  evolution of molecular clouds: a very narrow lognormal regime
  appears when the cloud is being assembled.  However, as the global
  gravitational contraction occurs, the initial density fluctuations
  are enhanced, resulting, first, in a wider lognormal \Npdf, and
  later, in a power-law \Npdf.  We thus suggest that the observed
  \Npdf\ of molecular clouds are a manifestation of their global
  gravitationally contracting state. We also show that, contrary to
  recent suggestions, the exact value of the power-law slope is not
  unique, as it depends on the projection in which the cloud is being
  observed.}


\end{abstract}


\section{Introduction}\label{sec:intro}

It has been known since the first observations of molecular gas in
star-forming clouds that the CO lines exhibit supersonic line widths
\citep{Wilson_etal70}.  \citet{Goldreich_Kwan74} suggested that such
supersonic linewidths could be produced by large-scale collapse of the
molecular clouds.  In contrast, \citet{Zuckerman_Evans74} dismissed
the idea of large-scale collapse with the argument that, if clouds
were collapsing freely, the star formation rate in the Galaxy should
be a factor of 100 times larger than observed, leading to quick
exhaustion of its gas content.  Those authors proposed instead that
the large linewidths are produced by small-scale supersonic
turbulence\footnote{Strictly speaking, \citet{Zuckerman_Evans74} ruled
out only {\em radial} large-scale collapse, which is hardly
surprising, as molecular clouds are generally far from round.}.

Since then, the scenario of supersonic molecular cloud turbulence has
received much attention, and turbulent models of molecular clouds have
been developed by several groups \citep[see][and references
  therein]{MK04, BP_etal07}.  However, the turbulent picture faces
several problems, not the least of which is that supersonic shocks
dissipate energy quickly, within one dynamical time scale, even if
turbulence is magnetized \citep{Stone_etal98, MacLow_etal98,
  Padoan_Nordlund99}. Moreover, although stellar energy has been
proposed as a source of turbulence within molecular clouds, it is not
clear that such feedback can maintain the identity of the clouds;
high-mass stars can easily blow apart clouds, and bipolar flows from
low-mass stars are so focused that general support seems unlikely
\citep[see review by][]{VS10a}.  Thus it is far from clear - and it
has not been demonstrated numerically - that stellar feedback can keep
clouds near equilibrium for several dynamical timescales.  In fact,
the available evidence suggest the opposite \citep[see ][]{VS_etal10}.

The idea that molecular clouds (MCs) are the result of large-scale
compressions of the H~I diffuse medium was explored with numerical
simulations of a 1~kpc squared piece of the galactic disk,
representative of the Solar Neighborhood by \citet{BVS99} and
\citet{BHV99}, and collected in a coherent scenario by
\citet{HBB01}. These authors found that MCs can be formed rapidly and
proceed to star formation almost at the same time, since the column
density threshold for gravitational instability and for molecular gas
formation are similar\footnote{Note, however, that the entire
evolutionary sequence that culminates with the formation of a giant
molecular cloud involves timescales of a few tens of Myr, of which the
``molecular stage'' may represent a small final fraction \citep{HBB01,
Bergin_etal04, VS_etal07, VS_etal10}}.  Simultaneously,
\cite{Hennebelle_Perault99} discussed the idea of large-scale
compressions of the diffuse medium undergoing thermal instability,
thus producing dense gas.  This led to the idea that molecular clouds
in the solar neighborhood are formed in large scale compressions from
the warm, diffuse, thermally unstable medium
\citep{Heitsch_etal06, VS_etal07, HH08}.  As these clouds accumulate
mass and cool rapidly, they become Jeans unstable; at this point,
gravity begins to dominate the motions, developing near equipartition
between the gravitational and kinetic energies \citep{VS_etal07,
  HH08}, erroneously but frequently inferred to be in ``Virial
Equilibrium'' \citep{BP06}.  As a consequence, the initial turbulent
fluctuations in density and velocity produce hierarchical
fragmentation, in which dense clumps with short free-fall times
proceed to a rapid collapse, while at the same time the whole cloud
contracts at a slower rate, because of its lower average density.

This model of molecular cloud formation has been subject to different
observational tests. For instance, \citet{HBH09} showed that the
CO(1-0) observed line profiles through such modeled clouds reproduce
the supersonic, turbulent line profiles of actual molecular clouds, as
well as the so-called core-to-core velocity dispersion.  In other
words, such global collapse of irregular structures develops internal
disordered, turbulent motions at some level, and thus the
large linewidths in actual MCs must be showing mainly the large-scale
systematic, inhomogeneous inward motions, rather than pure turbulence.
In fact, \citet{VS_etal09} showed that the culmination of the collapse
of a medium-sized cloud produces a dense clump with large velocity
dispersion, with physical properties similar to those of massive
star-forming regions. In this clump, the velocity field is dominated
by the collapse motions.

In addition, \citet{BP_etal11}, based on observations of massive dense
cores \citep{Caselli_Myers95, Plume_etal97, Shirley_etal03,
Gibson_etal09, Wu_etal10}, showed that the \citet{Larson81}
relationship between velocity dispersion and size arises naturally if
turbulent motions are the result of hierarchical and chaotic global +
local collapse.  In particular, they found that the velocity
dispersion should scale as

\begin{equation}
\delta v^2 \simeq 2 G \Sigma R
\end{equation}
where $\delta v$ is the velocity dispersion of the gas, $\Sigma$ is
the column density of the core, $R$ its size, and $G$ the universal
constant of gravity, according to the \citet{Heyer_etal09} relation
recently found for molecular clouds.

In the present work we take a further step to compare models with
observations. We analyze the column density probability distribution
function from models of molecular cloud formation and evolution via
warm, thermally unstable H~I stream collision.  The goal of the
present contribution is to assess the evolution of the column density
probability distribution function (\Npdf) in these models, and compare
it to the \Npdf\ of observed molecular clouds. This is particularly
important because a number of recent observations have reported
\Npdf\ for several nearby molecular clouds, permitting more tests of
theory.

In section \S\ref{sec:knowledge} we discuss the current knowledge of
volumetric and column density pdfs for interstellar gas. In
\S\ref{sec:models} we briefly describe the models used, and in
\S\ref{sec:results} we present the time evolution of our \Npdf s.
In \S\ref{sec:discussion} we present a brief discussion and in
\S\ref{sec:conclusions} we give our main conclusions.

\section{Volumetric and Column Density PDFs of molecular 
  clouds}\label{sec:knowledge}

One important tool for understanding the internal structure of
molecular clouds is the volume density probability distribution
(\rhopdf).  For isothermal, turbulent flows of a given Mach number
$M$, where shocks occur stochastically and in succession, creating
density fluctuations $\rho_1/\rho_0 \propto M^2$, the \rhopdf\ is a
lognormal function, i.e., a Gaussian function in the logarithm of the
density. \citet{VS94} and \cite{PV98} explain this result as follows:
in an isothermal flow, the speed of sound is spatially uniform, and a
shock of intensity $M$ will induce a density jump $\rho_1/\rho_0$ over
the mean density $\rho_0$.  If another shock of intensity $M$ arrives
at the place where the density is now $\rho_1$, it will induce another
density jump of size $\rho_2/\rho_1 \propto M^2$. Thus, if the flow is
stochastic, such density jumps must be spatially uniform, and a given
density distribution must be obtained by a succession of
multiplicative density jumps, which are additive in the logarithm.
Thus, at a given position in space, each density jump is independent
of the previous and the following ones, and therefore, the central
limit theorem, according to which the distribution of the sum of
identically-distributed, independent events approaches a Gaussian, can
be applied to the logarithm of the density, resulting in a lognormal
density probability distribution function.


Various authors suggested that the column density of turbulent,
isothermal flows must also be lognormal \citep[e.g.,
][]{Ostriker_etal01, Ridge_etal06}. However, this is only a
consequence of the fact that interstellar turbulence contains the
largest velocities at the largest scales, and thus, the size of the
interstellar clouds is comparable to their correlation length.  In
this case, \citet{VS_Garcia01} showed that the local values of the
volume density along a single line of sight are correlated, and thus
the column density might be representative of the mean value of the
volume density along such line of sight.  Thus, for different lines of
sight, the values of the \Npdf\ are correlated to the mean values of
the \rhopdf, and thus, the \Npdf\ follows the shape of the \rhopdf.
This would not be the case, however, if the correlation lengths of
molecular clouds were small. In this case, the line of sight will
contain many correlation lengths, and thus, the mean value will not be
representative of the volumetric density of the line of sight. In this
case, the central limit applies over the sum of several correlation
lengths, and the column density should exhibit a Gaussian shape
\citep{VS_Garcia01}.

In recent years, a number of observational studies have focused on
understanding the column density PDF of nearby clouds.  For instance,
\cite{Ridge_etal06} used extinction data from the COMPLETE
(COordinated Molecular Probe Line Extinction Thermal Emission) Survey
of Star-Forming Regions, to argue that the gas in Ophiuchus and
Perseus molecular clouds exhibits a lognormal \Npdf \citep[see
also][]{Goodman_etal09}. Unfortunately, the column density in this
work is plotted in linear scale in the $y$ axis (number of events),
and thus, the excess in the high column density part of the \Npdf\
cannot be properly appreciated. However, that excess clearly indicates
that some degree of departure of the lognormality is going on in
Perseus.

On the other hand, \citet{Froebrich_etal07} presented maps and \Npdf s
of 14 different regions inferred from extinction measurements. They
found that the \Npdf s exhibit different shapes, although the
dynamical range in visual extinction $A_V$ is frequently smaller than
one order of magnitude, making it difficult to draw firm conclusions
about the actual shape of the \Npdf\ for these clouds.

\citet{Kainulainen_etal09} took this a step further, showing that the
Coalsack and Lupus~V molecular clouds, which do not have signs of star
formation, exhibit lognormal \Npdf s, while clouds that have already
formed stars, like Taurus and Lupus~I, do exhibit power-law like tails
at large column densities .  Their results were confirmed almost
immediately afterwards by \citet{Froebrich_Rowles10}.  The standard
interpretation of both groups is that clouds that are turbulent and
non-star forming, exhibit a lognormal \Npdf, while clouds that are
forming stars are somehow decoupled, and exhibit a power-law tail at
large column densities.

The above observational results have a clear counterpart in numerical
simulations of star formation in turbulent, isothermal MCs. Several
groups have shown that the \rhopdf s in those simulations develop
power-law tails at late stages \citep{Klessen00, DB05, VS_etal08}. 
More recently, \citet{Kritsuk_etal10} have shown that centrally-peaked
density distributions, such as the singular isothermal sphere
\citet{Shu77}, or other dynamical solutions such as a pressure-free or
inside-out collapse, which are all characterized by power-law density
profiles, have \rhopdf s and \Npdf s with power-law high-density
tails, whose slope does in fact depend on the slope of the density
profile.  Also, they show that numerical simulations of
self-gravitating isothermal turbulence develop power-law tails as time
progresses.  Thus, they attribute the development of such tails in
self-gravitating turbulent flows to the formation of local collapsing
sites. Although they do not mention it explicitly, one can conjecture
that the PDFs of the entire turbulent flow exhibit the characteristic
power-law tails of the collapsing centers, because the latter are the
only sites where such high densities develop.  Note also that, because
those authors only evolved their simulations to less than half a
free-fall time, the dominant component of the kinetic energy in the
simulations was still that due to the turbulent flow.

On the other hand, multiphase simulations of the formation and
evolution of MCs out of the convergence of warm, diffuse gas in the
presence of self-gravity by \citet{VS_etal07, VS_etal10} and
\citet{HH08} show that the dominant component of the kinetic energy in
the clouds after they become dominated by self-gravity is generalized
gravitational contraction. In the remainder of this paper, we show that
those multiphase simulations also develop power-law tails in their
\Npdf s at late times, suggesting that the PDFs of MCs should not be
expected to be stationary, but rather to evolve in time, as a
consequence of the transition of the clouds from being
turbulence dominated to being collapse dominated.

As an alternative interpretation, \citet{Tassis_etal10} have recently
shown that a lognormal column density PDF does not necessarily imply
that the flow is turbulent. In particular, they have shown that
simulations of individual and multiple collapses also exhibit
lognormal \Npdf s at early stages in their evolution. However, it is
well known that isothermal compressible turbulence develops such a
lognormal PDF \citet{VS94, PV98}, Thus, it is natural to ascribe the
lognormal parts of PDFs of ISM turbulence simulations to supersonic
isothermal turbulence. Similarly, the high-density gas in the ISM is
characterized by supersonic linewidths and behaves very close to
isothermally, suggesting that lognormal PDFs in MCs are due to this
process as well.

In the next section we will show that this behavior can be understood
in a unified model of molecular cloud formation and evolution, where
clouds are assembled by large-scale convergent flows in the WNM, which
initially produce turbulence, but gravity rapidly takes over,producing
a hierarchy of nested collapsing motions, and thus, the \Npdf\
develops a power-law wing at large column densities.

\section{Results}\label{sec:results}

\subsection{Brief description of the models}\label{sec:models}

In the present section we analyze the evolution of the \Npdf\ in
numerical simulations of the formation and evolution of molecular
clouds.  In order to understand the evolution of the \Npdf, we briefly
describe the evolution of the clouds of those simulations \citep[for a
review, see ][]{VS10b}.

The models consist in the collision of two large streams of warm
thermally bi-stable H~I gas.  The streams collide with a given inflow
velocity, on top of which fluctuations in magnitude and shape are
superposed.

As the streams collide, the compressed regions undergo a nonlinearly
induced thermal instability, cooling down rapidly and becoming
turbulent due to a combination of various instabilities
\citep{Heitsch_etal05, Heitsch_etal06, VS_etal06}. As the cold clouds
accumulate mass, they quickly become Jeans unstable at various scales.
The superposition of those different centers of collapse produces
hierarchy of nested collapses, which is reflected in highly supersonic
line profiles \citep{HBH09, BP_etal11} that traditionally have been
interpreted as evidence of turbulence.

Under this general scheme, several studies \citep[e.g.,][]{VS_etal07,
HH08, VS_etal10} have performed simulations with different numerical
schemes (SPH, fixed grid or AMR, respectively), and with slightly
different initial conditions.  However, in spite of the differences,
the formation and evolution of the modeled molecular clouds exhibit
similar characteristics: initially, all modeled clouds fragment
rapidly due to dynamical an thermal instabilities, as well as the
initial velocity fluctuations.  However, while the models without
gravity remain fragmented and do not develop large density regions,
the models with gravity start to collapse transversely to the inflow
direction, developing local centers of collapse with large volumetric
densities ($\geq$~\diezala6~cm\alamenos 3), as well as large column
densities, ($\geq$~\diezala{24}~cm\alamenos 2).

As \citet{HH08} discuss, the same cloud at different evolutionary times
appears as if it were different kind of objects: at earlier times, the
compressed layer could be classified as a diffuse H~I cloud, with low
column densities and thus, very little CO content
\citep[see][]{Heitsch_etal06, VS_etal06}.  As time goes by, the layer
accumulates more mass, column densities increase, and some CO begins to
form.  At this stage the cloud could be identified as a
``translucent'' cloud.  Later in time, larger column densities are
achieved not only because of the compression of gas along the
direction of the streams, but because the transverse collapse causes a
rapid increase of the column density, shielding the interior of the
cloud against UV photons. At this stage, the cloud appears mostly
molecular, with dense cores that are already collapsing and which a
few Myr later form stars. The precise times in which each one of these
stages occurs depends on several parameters, e.g., the initial values
of the density and velocity fields, the actual values of their
fluctuations, the total mass of the streams, etc.  However, once the
cloud has accumulated enough mass to allow for molecule formation, the
first collapsed objects appear a few Myr later.

%
In the present work, in order to focus on the shape of the
\Npdf\ during the process of cloud formation, we make use of runs
without stellar feedback. We use runs HF and GF2 of the suite
presented by \citet{HH08} and LAF0 run by \citet{VS_etal10}. Run HF is
similar to GF2, but without self-gravity, and is used only for
reference.  Run GF2 represents the evolution of a $\sim 22\times
44\times 44$~pc cube filled with gas at $T\sim 1800$~K and at a
density of $n=3$~cm\alamenos 3. The incoming streams are ellipsoidal
cylinders with a cross section of 22~pc, an inflow velocity of
7.9~km/sec, and inflow boundary conditions on the numerical box, i.e.,
infinite incoming cylinders.  In run LAF0, the simulation is performed
using an adaptive mesh refinement numerical scheme \citep[ART code
  by][]{Kravtsov_etal97}. The box has a size of $256$ pc per side,
filled with a gas at $T=5000$~K, and a density of $n=1$~cm\alamenos 3.
The streams are circular cylinders having a diameter of 32~pc, and an
incoming velocity of 5.9~km/sec, and a length of 110~pc.  The inflows
are thus fully contained within the box. {This means not only that
  they have a finite length and duration, but also that the total mass
  of the box in the \citet{VS_etal10} runs is constant, while in the
  runs by \cite{HH08} it is not.  This result will have implications
  on the height of the \Npdf, as we will see in
  \S\ref{sec:columndenistypdfs}}.

In order to trigger instabilities in the compressed layer, in the HF
and GF2 runs the shock front has a sinusoidal shape, while in LAF0
run, a fluctuating velocity field is added to the inflow velocity with
rms amplitude of $\sim$~30\% of the inflow speed, and with $\sim$~1/2
the diameter of the inflows.  For more details, see conditions of runs
GF2 and LAF0 for \citet{HH08} and \citet{VS_etal10}, respectively.

\subsection{Column density PDFs}\label{sec:columndenistypdfs}

\begin{figure*}
\includegraphics[width=0.49\hsize]{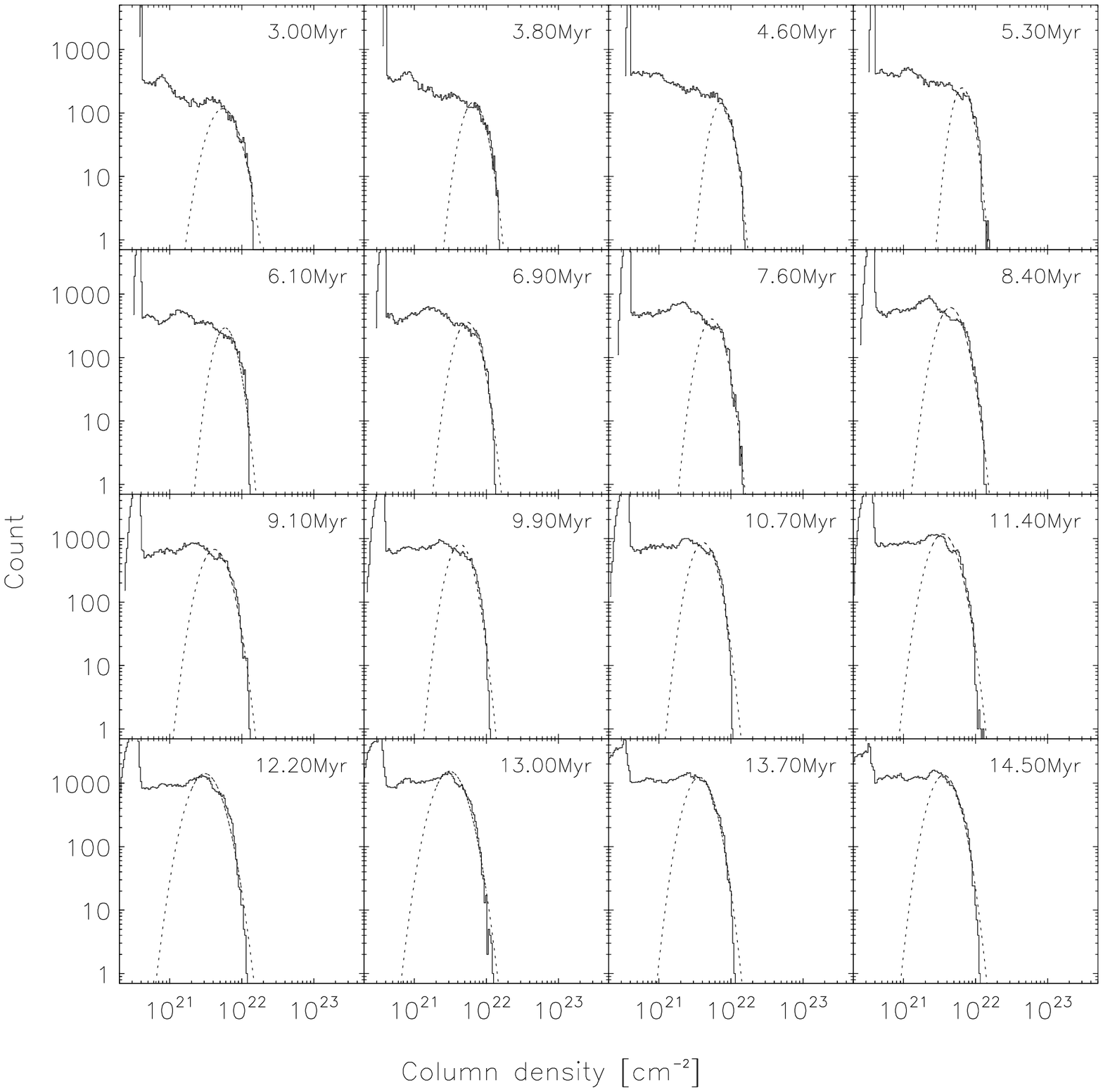}
\includegraphics[width=0.49\hsize]{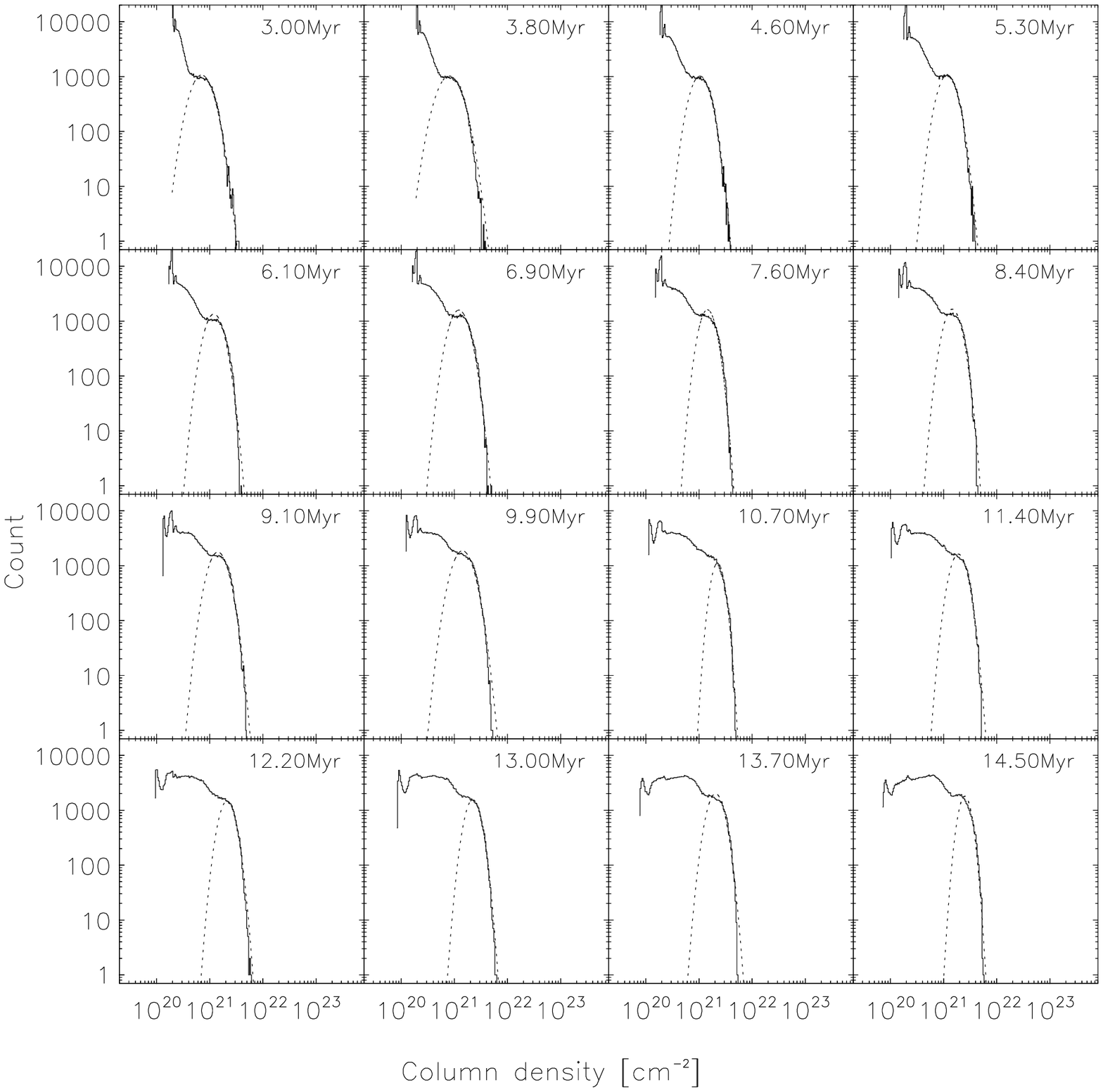}
\caption{\label{fig:Npdf-run_HF} a) Time evolution of the column
  density probability distribution function of a run without
  self-gravity \citep[run HF in][]{HH08} seen perpendicular to the
  flow (edge on). b) Same as a), but for the projection along the flow
  (face on).}
\end{figure*}

\begin{figure*}
\includegraphics[width=0.49\hsize]{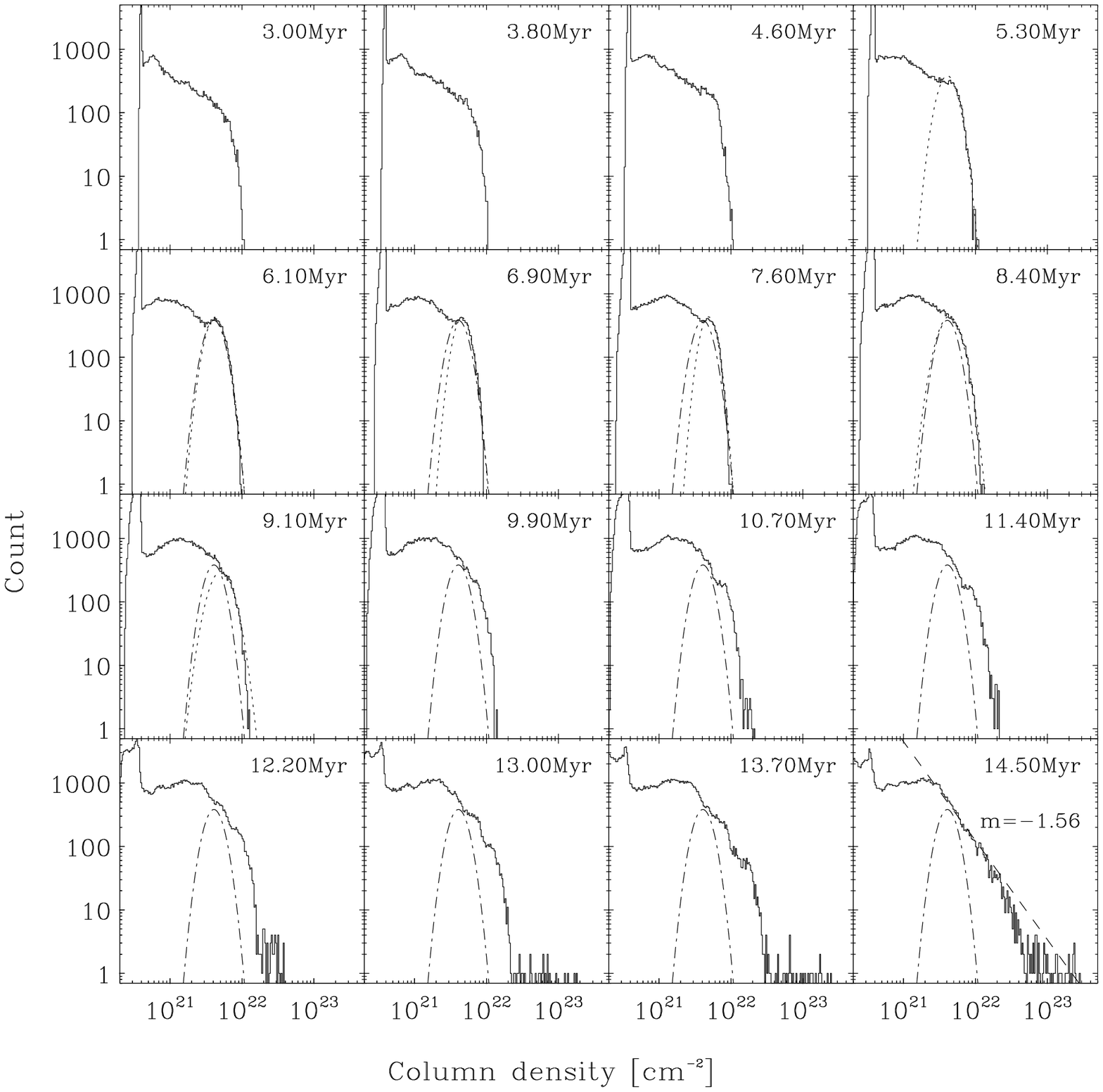}
\includegraphics[width=0.49\hsize]{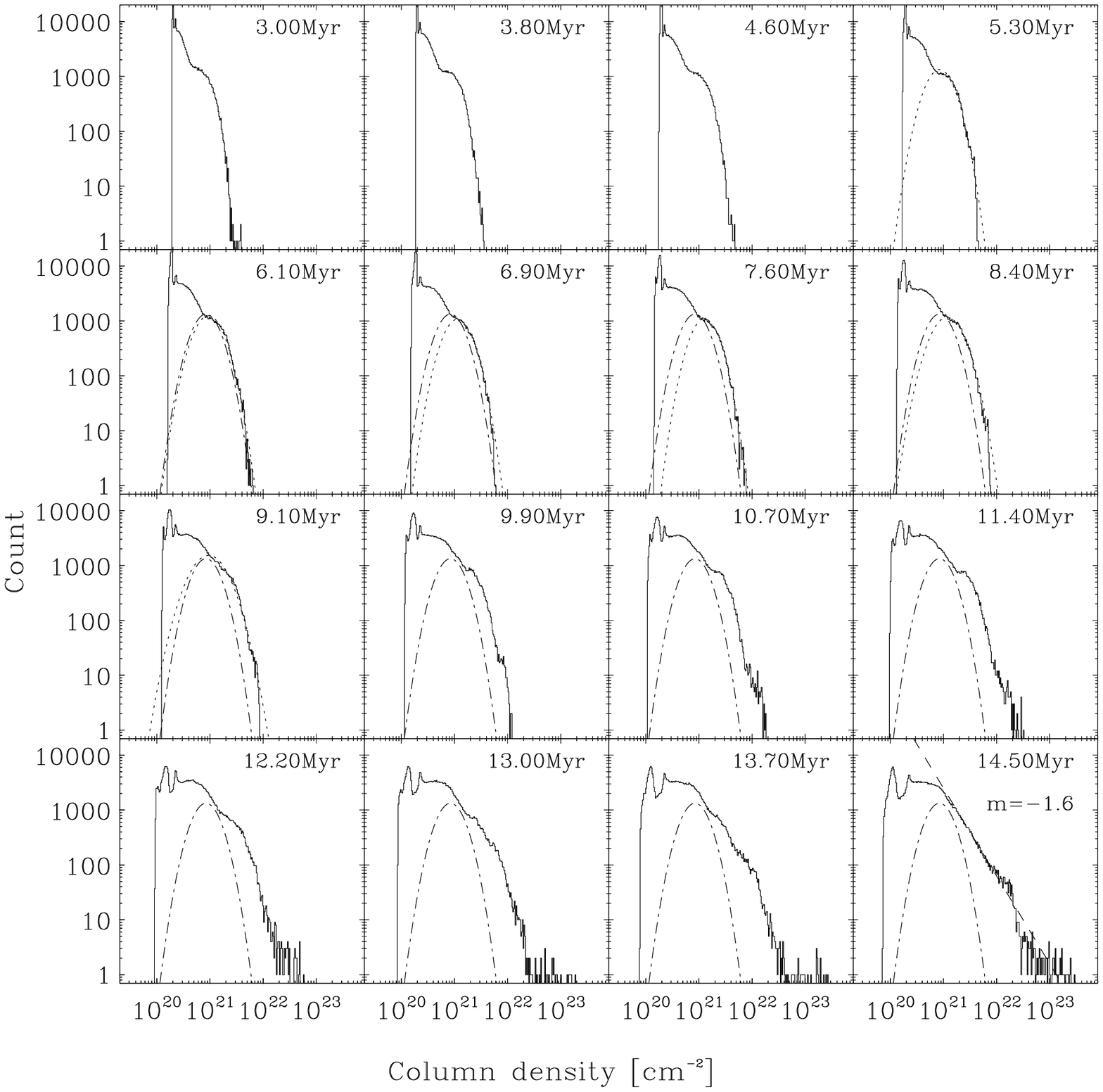}
\caption{\label{fig:Npdf-HH08}  a) Time evolution of the column
  density probability distribution function of a run including
  self-gravity \citep[run GF2 in][]{HH08} seen perpendicular to the
  flow (edge on). b) Same as a), but for the projection along the flow
  (face on). }
\end{figure*}

\begin{figure*}
\includegraphics[width=0.49\hsize]{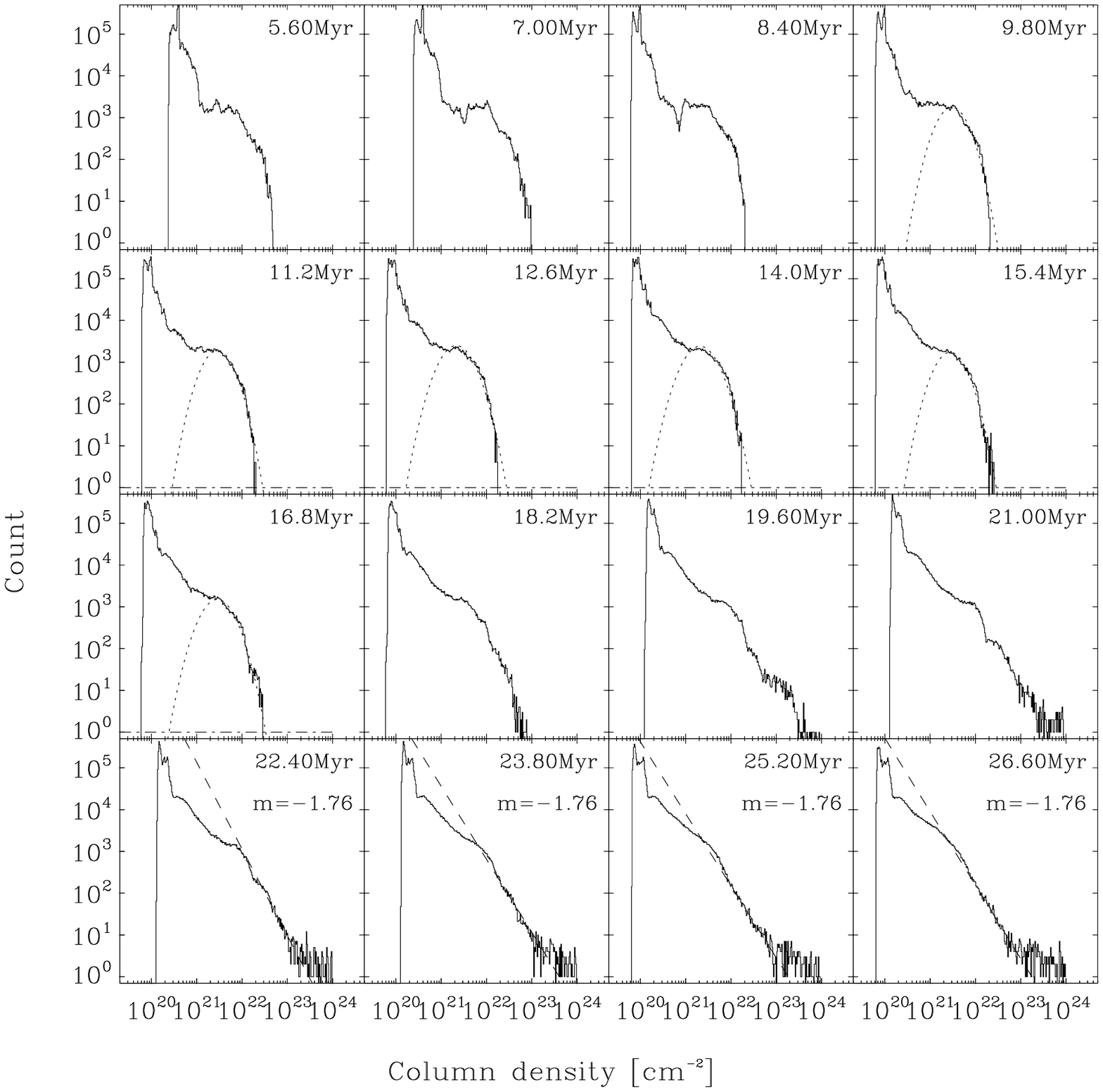}
\includegraphics[width=0.49\hsize]{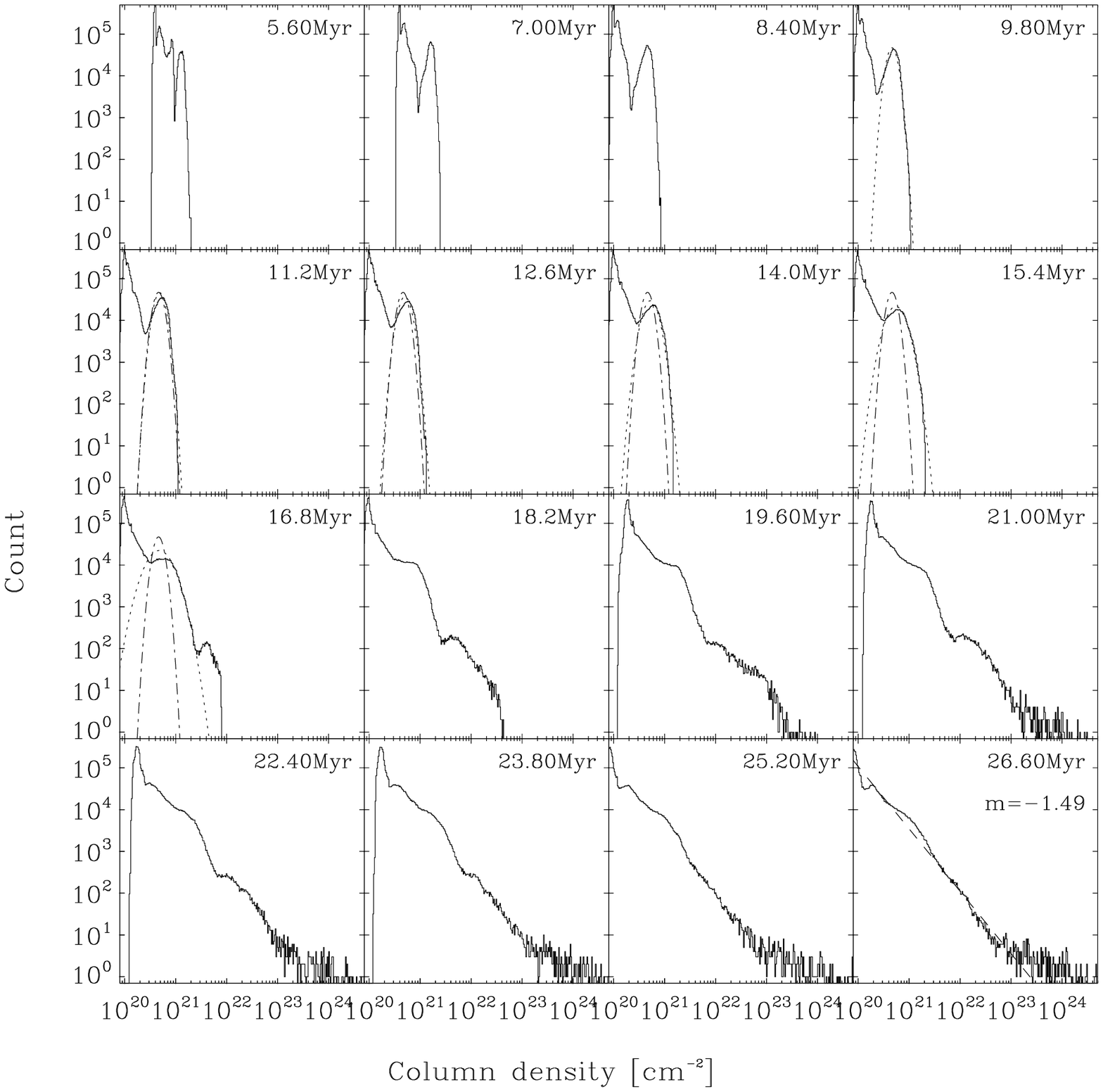}
\caption{\label{fig:Npdf-VS} a) Time evolution of the column density
  probability distribution function of model LAF0 in \citet{VS_etal10}
  seen perpendicular to the flow (edge on). b) Same as a), but for the
  projection along the flow (face on).}
\end{figure*}

{Figs.~\ref{fig:Npdf-run_HF}, \ref{fig:Npdf-HH08}, and
  \ref{fig:Npdf-VS} show the \Npdf s of models HF, GF2 and LAF0,
  respectively.  For each run, we present two extreme cases: \Npdf s
  calculated by projecting the density field perpendicular to the
  inflow and one projecting along it.  All plots exhibit 16 panels,
  from 3 to 15~Myr in Figs.\ \ref{fig:Npdf-run_HF} and
  \ref{fig:Npdf-HH08}, and from 5.6~Myr to 26.6~Myr in
  Fig.\ \ref{fig:Npdf-VS}. In both cases, the time is measured from
  the beginning of the collision.  The first point to notice in all
  three figures is that in all the simulations, at every time step,
  and in both projections, there is a large number of data points at
  low column densities.  This is just a consequence of the setup of
  the simulations: the incoming flows are cylindrical, and outside
  these cylinders the gas is uniform, with low volume density and zero
  initial velocity.  The peaks in each case simply reflect the column
  density of the uniform background medium.  The mean density of this
  field is $\sim$~1~cm\alamenos 3, and thus, the typical column
  density of this part of the box is $\sim 2\times
  10^{20}$~cm\alamenos 2 for run GF2 and $\sim 8\times
  10^{20}$~cm\alamenos 2 for the LAF0 one.

The second point to notice is that the column densities in both
projections are extreme cases because these are extreme
projections. Typically, molecular clouds must be observed at
intermediate angles.  In any case, the initial \Npdf s for the
projection along the flow (plane $y-z$) are narrower than the
projections perpendicular to the flow (planes $x-y$ or $x-z$, not
showed here). This is because the clouds are formed in the $y-z$
plane, and hence the cloud is thinnest along the $x$-direction.

In all three figures we further show lognormal and/or power-law fits
as follows: Fig.~\ref{fig:Npdf-run_HF} shows only lognormal fits
(dotted line) at every timestep, since the corresponding run is purely
hydrodynamical.  Note that the PDF increases in height as time goes
on. This is a consequence of the inflow boundary conditions of the
models from \citet{HH08}, which imply that the total mass in the box
increases in time.

In Figs.~\ref{fig:Npdf-HH08} and \ref{fig:Npdf-VS}, on the other
hand, we show the lognormal fits from $t=$~5.3~Myr to $t=$~9.1~Myr and
from $t=$~9.8 Myr to $t=$~16.8 Myr, respectively. The final times of
these intervals correspond to
the time at which the cloud start collapsing more vigorously. However,
for reference, we keep the first fit until the end
of the simulation (dot-dashed line). 

}

{We note that in both panels of Fig.~\ref{fig:Npdf-run_HF} the \Npdf s
  exhibit a lognormal-like wing at large column densities (dotted
  line) since early on in the simulation's evolution, and maintain this
shape for the rest of the evolution.  The 
  lognormal wing at large column densities is due to the density
  fluctuations} produced by the turbulence induced by a
  combination of nonlinear thin shell, thermal and Kelvin-Helmholtz
  instabilities in the compressed layer \citep{Heitsch_etal05}.
However, it is worth noting 
  that this lognormal part is quite narrow, with column densities
  spanning less than two orders of magnitude, from $\sim 2\times
  10^{21}$~cm\alamenos 2 to 10\ala{23}~cm\alamenos 2, for the entire
  duration of the evolution. This is due to the fact that the turbulence
generated by the instabilities is not very strong, in agreement with
previous results \citep{KI02}.

\begin{figure}
\includegraphics[width=\hsize]{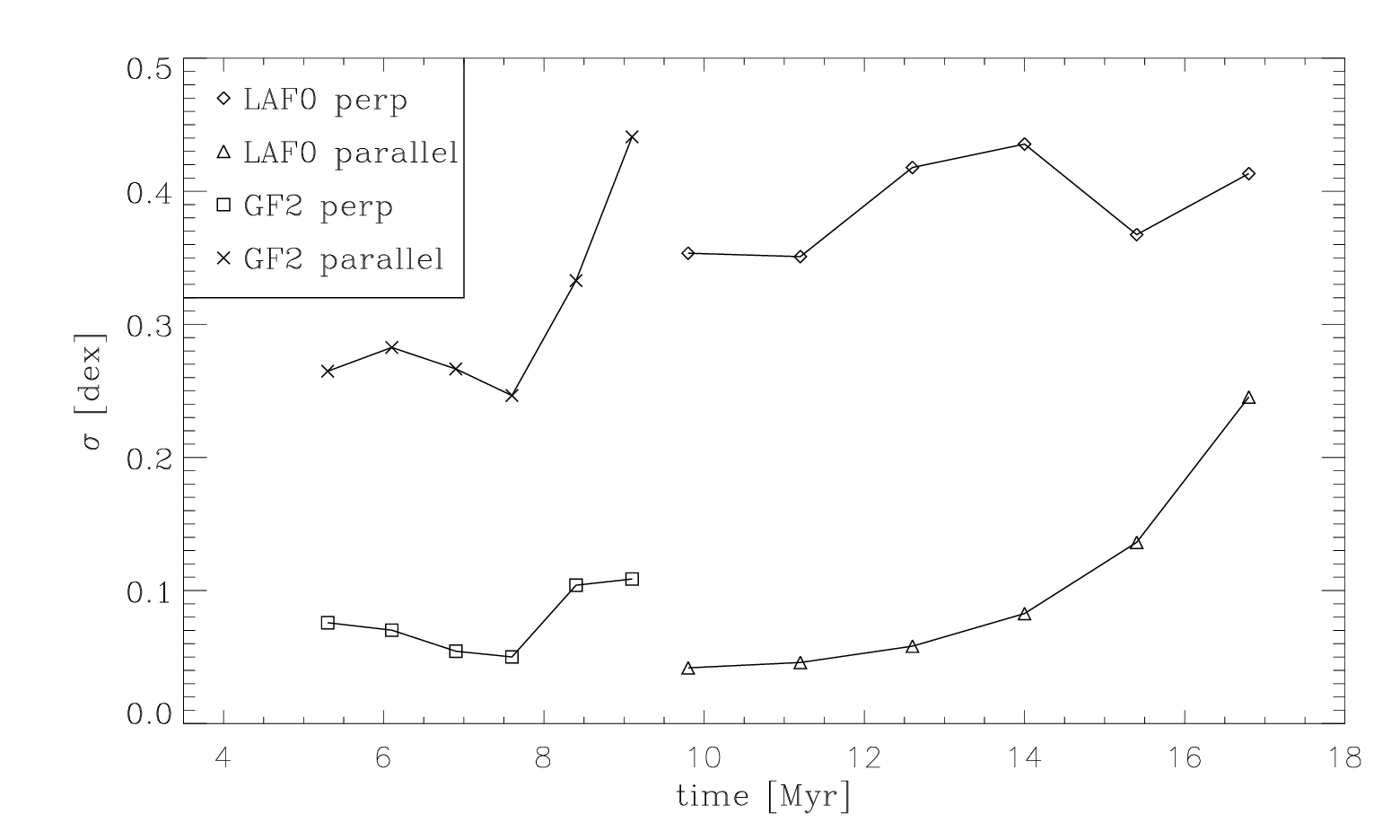}
\caption{\label{fig:sigma} {Time evolution of the standard
  deviation of the lognormal function fitted in the two projections of
  the two runs with self-gravity. As can be seen, in all cases
  $\sigma$ increases with time, implying that gravity makes wider the
  lognormal of a purely turbulent gas. }}
\end{figure}

{The case with self-gravity is quite different (see
  Figs.~\ref{fig:Npdf-HH08} and \ref{fig:Npdf-VS}).  We divide the
  behavior in two cases: the lognormal and the power-law behaviors.
  We first note that, after some transients, the dense column density
  region develops a narrow lognormal shape, indicated by the dotted
  line (which we mantain through all frames for clarity).  However, as
  time goes on, gravity modifies the lognormal shape, first by making
  it wider.  In order to show this widening more clearly, in
  Fig.\ref{fig:sigma} we plot the evolution of the standard deviation
  $\sigma$ of the lognormal fits as a function of time for the four
  runs with self-gravity. From this figure it is clear that the
  lognormals tend to increase in width as the collapse proceeds. This
  behavior has been recently reported by \citet{Tassis_etal10}.  }
Finally, at late times, gravity runs away and column density grows
several orders of magnitude in a few Myr. As a consequence, the
\Npdf\ develops a power-law tail at large column densities.

{The \Npdf s shown in Figs.~\ref{fig:Npdf-HH08} and
  \ref{fig:Npdf-VS} are reminiscent of the variety of \Npdf s for
  different clouds shown by \citet{Kainulainen_etal11}.  In fact, as
  shown by \citet{Froebrich_etal07}, every cloud may have its own
  slope, depending not only on the physical conditions of the cloud,
  but also on the resolution at which the cloud itself has been
  observed.  }

{Furthermore, by comparing the direction of the projections, we
  note that the slope depends on the orientation of the cloud:
  although in models GF2 the slope in both projections is almost the
  same (see Fig~\ref{fig:Npdf-HH08}), models LAF0
  (Fig~\ref{fig:Npdf-VS}) show that the slope is steeper when we
  observe the cloud along its longer dimension.  This effect occurs
  because if the cloud is mostly aligned with the line of sight, the
  observed column densities will be larger than the column densities
  obtained if the cloud is mostly on the plane of the sky.  This result
  contrasts with the suggestion by \cite{Kritsuk_etal10} that all the
  \Npdf s should tend to evolve towards a single value.  These authors
  used a model of a spherical cloud, and thus every projection is the
  same. However, real clouds are far from round, and thus one cannot
  expect an asymptotic single value of the slope of the power-law tail
  of the \Npdf.  }

Also, it is noteworthy that the other run with slightly
different perturbations in \citet{HH08}, (run GF1, not shown), which has a
smaller perturbation in the shock front, has an \Npdf\ evolution which
is basically indistinguishable from the one presented here for model GF2.
Similarly, the run with smaller amplitude fluctuations (SAF0, also not
shown) in 
\citet{VS_etal10}, follows an \Npdf\ evolution indistinguishable from
model LAF0.  However, the precise time in which each model
\citep{HH08, VS_etal10} start collapsing are different: $t\sim 10$~Myr
for the \citet{HH08} runs, and $t\sim 15$~Myr for the
\citet{VS_etal10} runs.  The difference seems to be due to the
free-fall time in each run: the compressed region in the \citet{HH08}
runs has 47\% of the volume of the compressed region in
\citet{VS_etal10}, but its mass flux is 5 times larger. Thus, the
density in the compressed region is 2.23 times larger in the
\citet{HH08} run, which in turn implies a free-fall timescale 2/3
times smaller.  This result suggests that, to first order, the main
parameters that define the time of collapse are the velocity and the
density of the incoming field.  Indeed, a parameter study by
\citet{Rosas-Guevara_etal10} has found widely different star formation
efficiencies and times of the onset of star formation as parameters
such as the inflow velocity and the amplitude of the background
turbulent fluctuations are varied.

%
%



In both runs, after the development of the lognormal \Npdf, a clear
power-law tail develops at high column densities.  This shape
transition marks also the beginning of the accelerated large-scale
collapse ongoing simultaneously with the small-scale local collapses,
and occurs in few Myr ($\sim 4-5$~Myr), regardless of the particular
setup of the runs.  It seems thus that once the cloud has achieved
enough shielding to form molecules, the cloud proceeds to collapse in
few Myr, as pointed out by \cite{HH08}.  The rapidity of this
transition explains why most of the molecular clouds in the Solar
Neighborhood exhibit young star forming regions (few Myrs), and there
are only few clouds with no signs of star formation \citep[Coalsack,
  Lupus~V, see ][]{Kainulainen_etal09}.

Thus, we conclude that as global$+$local collapses proceed, the number
of high column density points in the \Npdf s increases, causing a high
 \Npdf\ tail. 
As pointed out by \citet{Franco_Cox86, HBB01} and \citet{HH08}, in the
Solar Neighborhood the pressure due to self-gravity becomes importat
at about the
column density necessary to shield the compressed region against UV
radiation (\diezala{21}~cm\alamenos 2), allowing the formation of
molecules. This situation corresponds to the 3rd row in the rightmost
column of Fig.~1 in \citet{HH08}, and corresponds to the moment in
which the cloud has already started to collapse in the direction
perpendicular to the shock. At this moment, the densest cores of the
cloud reach column densities of \diezala{23}~cm\alamenos 2.


\section{Discussion}\label{sec:discussion}

\citet{Froebrich_etal07}, \citet{Kainulainen_etal09} and
\citet{Froebrich_Rowles10}, argue that the different shapes of the
observed \Npdf s must be due to a change in the governing physical
processes in molecular clouds. In particular,
\citet{Froebrich_Rowles10} argue that most of their \Npdf s can be
fitted by a lognormal distribution, plus a power-law wing, the first
one for low column densities, which corresponds to the gas that is not
forming stars, and the second one for the gas which is forming stars.

{Our results show that these two regimes (quiescent vs. active
  star formation) may very well be part of a single evolutionary
  process, as depicted by \citet{Kainulainen_etal09}. }  In the early
  stages of the molecular cloud, the H~I colliding streams form the
  molecular gas and provide part of the supersonic turbulent kinetic
  energy observed in the line profiles.  During this stage, the \Npdf\
  is lognormal, and most of this stage occurs in the atomic phase
  \citep{VS_etal07, HH08}, during which the cloud does not form stars.
  Later on, hierarchical and chaotic gravitational contraction takes
  over, {tending to produce a power-law tail in the \Npdf s. The
initial manifestation of this effect is that the lognormal appears to
become wider.} 
It is important to mention
  that gravity is required in order to achieve the 1.5--2 orders of
  magnitude in the spread of the column density values, since pure
  turbulence produces only very narrow lognormal \Npdf s.  {The
  contribution of self-gravity in increasing the column density helps
  the cloud to achieve values high enough to rapidly form molecular
  gas, as pointed out by \citet{HH08, HBH09}. A few Myr after the
  molecules have formed, the densest cores collapse
  and the first stars begin to form.}

Although we agree with the general scheme depicted by
\citet{Kainulainen_etal09}, we stress that the contribution of
self-gravity is indispensable to achieve high values of the column
density.  That is, self-gravity is not only important in driving the
collapse of gravitationally unstable clumps, but also in {\it forming}
such clumps, which would not be produced by turbulence alone \citep{VS_etal08}.
During the later stages, the primary source of ``turbulent''
motions (i.e., supersonic linewidths) in the molecular cloud is
gravity as well, with local mass concentrations that render the
velocity field more locally complex \citep{HBH09, BP_etal11}.

\section{Conclusions}\label{sec:conclusions}

{Using numerical models of the formation and evolution of
  molecular cloud from the diffuse interstellar medium, we have
  confirmed the suggestion by \citet{Kainulainen_etal09} that} the
\Npdf\ transits from a lognormal shape at early times, when the
kinetic energy in the clouds is dominated by their initial turbulence
(produced by various instabilities in the compressed layer that
becomes the cloud), to having a power-law tail at high $N$ at late
times, when their kinetic energy is dominated by gravitational
contraction, which culminates with the formation of stars. Our results
thus explain naturally recent observations showing that
non-star-forming clouds exhibit lognormal \Npdf s, while star-forming
clouds exhibit power-law tails at high densities, understanding the
result as the consequence of the transition of the clouds from a more
diffuse, turbulence-dominated regime to a denser, star-forming,
collapsing one.  {We also have shown that the slope of the
  power-law tail produced by gravity in the \Npdf\ does not have a
  unique value, but depends on the orientation of the cloud with respect to
  the line of sight. }


\section*{Acknowledgements}
This work has received partial support from grants UNAM/DGAPA IN110409
to JBP, CONACYT 102488 to EVS, NSF AST-0807305 to LH and FH, and has
made extensive use of the NASA's Astrophysics Data System Abstract
Service.

\label{lastpage}

 \end{document}